\documentstyle[sprocl,epsfig]{article}

\include{epsfig}

\newcounter{saveeqn}%
\newcommand{\alpheqn}{\setcounter{saveeqn}{\value{equation}}%
\stepcounter{saveeqn}\setcounter{equation}{0}%
\renewcommand{\theequation}
{\mbox{\arabic{saveeqn}-\alph{equation}}}}%
\newcommand{\reseteqn}{\setcounter{equation}{\value{saveeqn}}%
\renewcommand{\theequation}{\arabic{equation}}}%

\def\Journal#1#2#3#4{{#1} {\bf #2}, #3 (#4)}

\def\NPA{{\em Nucl. Phys.} A}

\def\PRL{\em Phys. Rev. Lett.}

\def\PRE{{\em Phys. Rev.} E}

\def\be{\begin{equation}}
\def\ee{\end{equation}}
\def\bea{\begin{eqnarray}}
\def\eea{\end{eqnarray}}

\begin{document}

\title{NET-BARYON FLUCTUATIONS AT THE QCD CRITICAL POINT}

\author{N. G. ANTONIOU, F. K. DIAKONOS AND A. S. KAPOYANNIS}

\address{Department of Physics, University of Athens, GR-15771,\\
Athens, Greece}

\maketitle\abstracts{In the vicinity of the quark-hadron critical point, in
the phase diagram of QCD, simple power-law relations constrain the
mid-rapidity net-baryon density profile, for different heavy-ion processes,
in a unifying scheme. The corresponding scaling variable specifies the
proximity of a given experiment to the critical point. An
important phenomenological aspect of the critical properties in the baryonic
sector is the possibility to observe intermittency effects in the phase space
distribution of net baryons.}

\section{Introduction}

Recent investigations concerning the QCD phase diagram $(\rho~-~T)$ predict
the existence of an endpoint along the critical line of the first order
quark-hadron phase transition \cite{Wilc}. This critical point, located on a
line of nonzero baryonic density $\rho=\rho_c$, is of second order and
belongs to the universality class of a $3-d$ Ising system.
The QCD critical point communicates with the hadronic world through
the fluctuations of a scalar field ($\sigma$-field)
which carries the quantum numbers of an isoscalar ($\sigma$-meson)
as the manifestation of a quark condensate, $\sigma \sim \langle \bar{q}q
\rangle$, in thermal environment.
In the effective theory of the QCD critical point, the classical
$\sigma$-field is a natural order parameter, the fluctuations of which obey
scaling laws dictated by the critical exponents of the $3-d$ Ising system
($\eta \approx 0, \beta \approx \frac{1}{3}, \delta \approx 5, \nu \approx
\frac{2}{3}$).
In a baryonic environment, however, the chiral condensate is expected to
have, at $T=T_c$, a strong dependence on the net-baryon density, driving the
$\sigma$-field close to zero for $\rho \approx \rho_c$:
$\langle \bar{q} q \rangle_{\rho} \approx \lambda \left(
\frac{\rho - \rho_c}{\rho_c}\right) \langle \bar{q} q \rangle_{0} +
O[(\rho-\rho_c)^2]$
where $\lambda$ is a dimensionless constant of the order of unity.
This dependence suggests a new order parameter,
$m=\rho-\rho_c$, associated with the critical properties of the baryonic
fluid created
in a quark-hadron phase transition. In fact, approaching the critical
point in the phase diagram, both the $\sigma$-field
fluctuations and the fluctuations of the order parameter $m(\vec{x})$
obey the same scaling laws ($\langle \bar{q} q \rangle_{\rho} \sim
m(\vec{x})$).

In the next section we will exploit the scaling properties of the order
parameter $m(\vec{x})$, properly adjusted to measurable quantities,
in heavy ion collisions. For this purpose we consider net-baryon production
in collisions of heavy nuclei with total number of participants $A_t$
and initial energy corresponding to a total size in rapidity $\Delta y = L$.

\section{Scaling of the baryonic density}

The created baryons in the process of quark-hadron phase transition occupy
a cylindrical volume with transverse radius $R_{\perp} \sim A_{\perp}^{1/3}$
and longitudinal size $L$ (in rapidity). The parameter $A_{\perp}$
specifies the effective number of participants, contributing to the
transverse geometry of the collision, and it is assumed $A_{\perp} \approx
\frac{A_t}{2}$, valid both for central ($A_{\perp}=A_{min}$)
and non-central collisions. Projecting out the net-baryon system onto the
longitudinal direction we end up with a $1-d$ liquid confined in a finite
rapidity region of size $L$ with local density $\rho(y)=\frac{n_b(y)}
{\pi R_{\perp}^2 \tau_f}$,
directly related to the measurable net-baryon density in rapidity
$n_b(y)=\frac{dN_b}{dy}$. Putting $R_{\perp}=R_o A_{\perp}^{1/3}$, we
introduce a characteristic volume $V_o=\pi R_o^2 \tau_f$
in terms of the freeze-out time scale $\tau_f \geq 6-8~fm$ and the
nuclear-size scale $R_o$ which contains also any growth
effects near the critical point ($R_o \geq 1.2~fm$).                                     ).
Using $V_o^{-1}$ as a scale for baryonic, freeze-out densities, the order
parameter $m(y)$ of the $1-d$ baryonic liquid is written:
\begin{equation}
m(y)=A_{\perp}^{-2/3} n_b(y) - \rho_c~~~~~,~~~~~0 \leq y \leq L
\label{eq:eq1}
\end{equation}
Near the critical point $(T \to T_c)$ the order parameter $m(y)$ obeys
a scaling law:
\begin{equation}
m(y)=t^{\beta} \left[ F_o(y/L) + t F_1(y/L) + ... \right]
\label{eq:eq2}
\end{equation}
where $t \equiv \frac{T_c - T}{T_c}$ $(T \leq T_c)$ and $\beta$ is the
appropriate critical exponent ($\beta \approx 1/3$). The leading
term $F_o(y/L)$ in the series (\ref{eq:eq2}) is a universal scaling function
whereas the nonleading terms $F_i(y/L)$ $(i=1,2,..)$ are
nonuniversal but $A$-independent quantities. Using eqs.(\ref{eq:eq1}) and
(\ref{eq:eq2}) we may specify the freeze-out line in the phase diagram by
considering the measurable, bulk density at midrapidity $n_b=n_b(L/2)$ as a
function of the freeze-out temperature $T_f$:
\begin{equation}
A_{\perp}^{-2/3} n_b = \rho_c + t_f^{\beta} \left[ F_o + t_f F_1 +
...\right]
\label{eq:eq3}
\end{equation}
where $t_f=\frac{T_c - T_f}{T_c}$ and $F_i \equiv F_i(1/2)$.
Integrating now eq.(\ref{eq:eq2}) in the
interval $0 \leq y \leq L$ we obtain at $T=T_f$:
\begin{equation}
A_{\perp}^{-2/3} A_t L^{-1} = \rho_c + t_f^{\beta} (I_o + t_f I_1 + ...)
\label{eq:eq4}
\end{equation}
where $I_i=\int_0^1 F_i(\xi) d\xi$. Introducing the variable
$z_c=A_{\perp}^{-2/3} A_t L^{-1}$ we find from eqs.(\ref{eq:eq3}) and
(\ref{eq:eq4}) a scaling law for the net-baryon density at midrapidity:
\begin{equation}
A_{\perp}^{-2/3} n_b = \Psi(z_c,\rho_c)~~~~~;~~~~~z_c \geq \rho_c
\label{eq:eq5}
\end{equation}
where the scaling function $\Psi(z_c,\rho_c)$ has the property
$\Psi(z_c=\rho_c,\rho_c)=\rho_c$. In the crossover regime $z_c < \rho_c$,
where critical fluctuations disappear, the local density at midrapidity is,
to a good approximation, $n_b \approx A_t L^{-1}$ suggesting a
continuous extension of the scaling law (\ref{eq:eq5}) in this region
with $\Psi(z_c,\rho_c)=z_c$ ($z_c < \rho_c$). It is of interest to note that
although the scaling function $\Psi(z_c,\rho_c)$ is continuous at the
critical point $(z_c=\rho_c)$, the first derivative is expected to be
discontinuous at this point in accordance with the nature of the phase
transition (critical point of second order). Keeping only the next to
leading term $F_1(y/L)$ in (\ref{eq:eq2}) we finally obtain:
\alpheqn
\begin{eqnarray}
A_{\perp}^{-2/3} n_b &=& \rho_c +
\frac{F_o}{F_1} [f(z_c,\rho_c)]^{\beta} + \nonumber\\
&&+ C [f(z_c,\rho_c)]^{\beta+1}  \\
\label{eq:eq6a}
A_{\perp}^{-2/3} n_b &=& \rho_c + F_o t_f^{\beta} (1 +
t_f \frac{C I_o^{1+\frac{1}{\beta}}}{F_o}) \\
\label{eq:eq6b}
f(z_c,\rho_c) &=& \frac{1}{G}(-1 + \sqrt{1 +
2 G (z_c -\rho_c)^{1/\beta}}); \nonumber\\
&&z_c \geq \rho_c
\label{eq:eq6c}
\end{eqnarray}
\reseteqn
where $C=\frac{F_1}{I_o^{1+\frac{1}{\beta}}}$, $G=\frac{2 I_1}{\beta
I_o^{1+\frac{1}{\beta}}}$.

The component $F_o(y/L)$ which dominates the order parameter
$m(y)$ in the limit $T \to T_c$ is approximately constant in the central
region $(y \approx \frac{L}{2}~,~L \gg 1)$, far from the walls (at the points
$y=0,L$) due to the approximate translational invariance of the finite system
in this region. On the other hand, approaching the walls, $F_o(y/L)$
describes the density correlation with the endpoints and obeys appropriate
power laws.
The solution which fulfils these requirements is:
\begin{equation}
F_o(\xi)=g[\xi (1- \xi)]^{-\beta/\nu}~~(0 \leq \xi \leq 1) 
\label{eq:eq7}
\end{equation}
and the final form of the scaling function $\Psi(z_c,\rho_c)$ becomes:
\alpheqn
\begin{eqnarray}
\Psi(z_c,\rho_c)&=&\rho_c+\frac{4^{\beta/\nu}}{B(1-\frac{\beta}{\nu},
1-\frac{\beta}{\nu})} \left[f(z_c,\rho_c)\right]^{\beta} \nonumber \\
&&+ C \left[f(z_c,\rho_c)\right]^{\beta +1}~;~z_c \geq \rho_c \\
\label{eq:eq8a}
\Psi(z_c,\rho_c)&=& z_c~~;~~z_c < \rho_c
\label{eq:eq8b}
\end{eqnarray}
\reseteqn
where $f(z_c,\rho_c)$ is given by eq.(\ref{eq:eq6c}). It is straightforward to
show that the discontinuity of the first derivative of $\Psi(z_c,\rho_c)$ at
$z_c=\rho_c$ is a nonzero universal constant:
$disc\left(\frac{d \Psi}{d z_c}\right)_{\rho_c} \approx 1 - \frac{2}{\pi}$
($\frac{\beta}{\nu} \approx \frac{1}{2}$) as expected from the
characteristic properties of a second-order phase transition.

\section{Approaching the critical point in experiments with heavy ions}

Combining eqs.(\ref{eq:eq5}-8) one may propose a framework for
the treatment of certain phenomenological aspects of the QCD critical point.
The scaling law (\ref{eq:eq5}) combined with eq.(8) involves
three nonuniversal parameters: (a) the critical density $\rho_c$ and (b) the
constants $C,G$ which give a measure of the nonleading effects, allowing
to accomodate in the scaling function (8) processes not very
close to the critical point. We have used measurements at the SPS in order
to fix these parameters on the basis of eqs.(\ref{eq:eq5}) and
(8). More specifically, in a series of experiments (Pb+Pb, S+Au,
S+Ag, S+S) with central and noncentral (Pb+Pb) collisions at the SPS
net baryons have been measured at midrapidity whereas the
scaling variable $z_c = A_{\perp}^{-2/3} A_t L^{-1}$, associated with these
experiments, covers a sufficiently wide range of values $(1 \leq z_c \leq 2)$
allowing for a best fit solution. The outcome of the fit is consistent with
the choice $G \approx 0$ and the equations (6) are simplified as follows:
\alpheqn
\begin{eqnarray}
A_{\perp}^{-2/3} n_b &=& \rho_c + \frac{2}{\pi} (z_c-\rho_c)
+ C (z_c-\rho_c)^4 \\
\label{eq:eq9a}
A_{\perp}^{-2/3} n_b &=& \rho_c + \frac{2 I_o}{\pi} t_f^{1/3} (1 +
t_f \frac{\pi C I_o^{1/3}}{2}) \\
\label{eq:eq9b}
A_{\perp}^{-2/3} n_b(y) &=& \rho_c +
\frac{(z_c-\rho_c)L}{\pi \sqrt{y(L-y)}} \nonumber\\
&&+ O[(z_c - \rho_c)^4]
\label{eq:eq9c}
\end{eqnarray}
\reseteqn
In eqs.(9) we have used the approximate values of the critical
exponents $\beta \approx \frac{1}{3}$, $\frac{\beta}{\nu} \approx
\frac{1}{2}$ in the $3-d$ Ising universality class. We have also
added eq.(\ref{eq:eq9c}) which gives the universal behaviour of the
net-baryon density $n_b(y)$, in the vicinity of the critical point
$(z_c -\rho_c \ll 1)$. The fitted values of the parameters in
eq.(\ref{eq:eq9a}) are $\rho_c =0.81$, $C=0.68$
and the overall behaviour of the solution is shown in
Fig.~1. In turns out that the critical density is rather small, compared
to the normal nuclear density $\rho_o \approx 0.17~fm^{-3}$: $\rho_c \leq
\frac{\rho_o}{5}$ ($R_o \geq 1.2~fm$, $\tau_f \geq 6~fm$),
suggesting that the critical temperature remains close to the value
$T_c \approx 140~MeV$
obtained in studies of QCD on the lattice at zero chemical potential
\cite{Laer}.

\begin{figure}[htb]
\begin{center}
\mbox{\epsfig{file=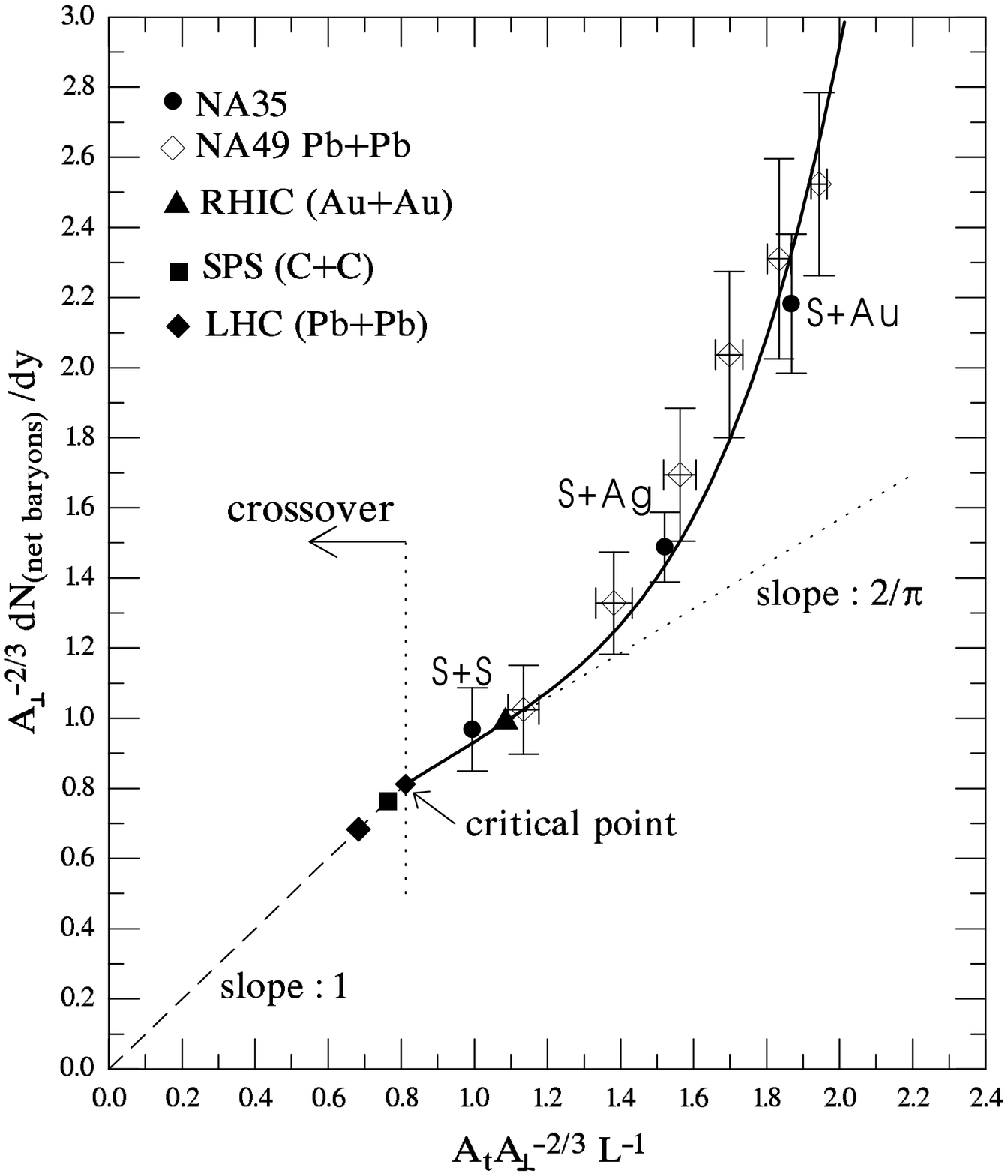,width=0.7\textwidth,angle=0}}
\caption{The scaling law (8) is illustrated together
with measurements at the SPS. The critical point and the corresponding
break in the slope of $\Psi(z_c,\rho_c)$ are also shown.}
\label{fig:fi1}
\end{center}
\end{figure}

The difference $d_c=\vert z_c -\rho_c \vert$ 
in Fig.~1 is a measure of the proximity of a given experiment $(A_t,L)$
to the critical point. We observe that the central Pb+Pb collisions at
the SPS ($d_c \approx 1.1$) drive the system into the most distant freeze-out
area, from the critical point, as compared to other processes at the same
energy. In fact, the most suitable experiments to bring quark matter close
to the critical point at the SPS are: S+S $(d_c \approx 0.18)$, S$_i$+S$_i$
$(d_c \approx 0.20)$ and C+C $(d_c \approx 0.06)$                          ),
central collisions.

\section{Critical fluctuations in the baryonic sector}

Once the deconfined phase of quark matter has reached the critical point in
a particular class of experiments, as discussed in the previous section,
strong critical fluctuations are expected to form intermittency patterns both
in the pion and net-baryon sector. As already mentioned in the introduction,
the origin of these fluctuations can be traced in the presence, at $T=T_c$,
of a zero mass field with a classical profile ($\sigma$-field) which, under
the assumption of a phase transition in local thermal equilibrium, is
described by an effective action in $3-d$, the projection of which onto
rapidity space is written as follows \cite{Anto1}:
\begin{equation}
\Gamma_c \approx \frac{\pi R_{\perp}^2}{C_A} \int_{\delta y} dy \left[
\frac{1}{2} \left( \frac{\partial \sigma}{\partial y} \right)^2 + 2 C_A^2
\beta_c^4 \sigma^{\delta +1} \right]~~~;~~~C_A=\frac{\tau_c}{\beta_c},~~~
\beta_c=T_c^{-1}
\label{eq:eq10}
\end{equation}
Equation (\ref{eq:eq10}) gives the free energy of the $\sigma$-field within
a cluster of size $\delta y$ in rapidity and $R_{\perp}$ in transverse space.
The critical fluctuations generated by (\ref{eq:eq12}) in the pion sector
have been studied extensively in our previous work \cite{Anto1}, therefore,
in what follows, we are going to discuss the fluctuations induced by the
$\sigma$-field in the net-baryon sector, noting that a direct measurement
of these fluctuations becomes feasible in current and future heavy-ion
experiments. For this purpose we introduce in eq.(\ref{eq:eq10}) the order
parameter $m(y)$ through the following equations:
\begin{eqnarray}
\sigma(y)\approx F \beta_c^2 m(y)~~~;~~~F \equiv -\frac{\lambda \langle
\bar{q}q \rangle_o}{\rho_c}~~~;~~~\langle \bar{q} q \rangle_o \approx
-3~fm^{-3} \nonumber \\
\Gamma_c \approx g_1 \int_{\delta y} dy \left[ \frac{1}{2} \left(
\frac{\partial \hat{m}}{\partial y} \right)^2 + g_2 \vert \hat{m}
\vert^{\delta +1} \right]~~~~~;~~~~~\hat{m}(y)=\beta_c^3 m(y) 
\label{eq:eq11}
\end{eqnarray}
where: $g_1=F^2\left(\frac{\pi R_{\perp}}{C_A \beta_c^2}\right)$,
$g_2=2 C_A^2 F^4$. The partition function $Z=\int {\cal{D}}[\hat{m}]
e^{-\Gamma_c[\hat{m}]}$ for each cluster is saturated by instanton-like
configurations \cite{Anto2} which for $\delta y \leq \delta_c$ lead to
self-similar structures, characterized by a pair-correlation function of the
form:
\begin{equation}
\langle \hat{m}(y) \hat{m}(0) \rangle \approx \frac{5}{6}
\frac{\Gamma(1/3)}{\Gamma(1/6)} \left(\frac{\pi R_{\perp}^2 C_A}{\beta_c^2}
\right) F^{-1} y ^{-\frac{1}{\delta +1}}
\label{eq:eq12}
\end{equation}
The size, in rapidity, of these fractal clusters is $\delta_c \approx
\left(\frac{\pi R_{\perp}^2}{16 \beta_c^2 C_A^2}\right)^{2/3}$ according to
the geometrical description of the critical systems \cite{Anto2}. Integrating
eq.(\ref{eq:eq12}) we find the fluctuation $\langle \delta n_b \rangle$ of
the net-baryon multiplicity with respect to the critical occupation number
within each cluster, as follows:
\begin{equation}
\langle \delta n_b \rangle \approx F^{-1}
\left(\frac{\pi R_{\perp}^2 C_A}{2 \beta_c^2}\right)
\frac{2^{2/3} \Gamma(1/3)}{\Gamma(1/6)} \delta_c^{5/6}
\label{eq:eq13}
\end{equation}
The dimensionless parameter $F$ is of the order $10^2$ and the size
$\delta_c$, on general grounds $(R_{\perp} \stackrel{<}{\sim} 2 \tau_c)$ is
of the order of one $(\delta_c \stackrel{<}{\sim} 1)$. As a result
the global baryonic system (in RHIC the
size of the system is $L \approx 11$) develops fluctuations at all scales
in rapidity since the direct correlation (\ref{eq:eq12}) propagates along the
entire system through the cooperation of many self-similar clusters of
relatively small size ($\delta_c \approx 0.35$ and $\langle \delta n_b
\rangle \approx 140$). We have quantified this mechanism in a Monte-Carlo
simulation for the conditions of the experiments at RHIC ($L \approx 11$) in
order to generate baryons with critical fluctuations. The distribution of
these ``critical" baryons in the rapidity space for a typical event as well
as the corresponding intermittency analysis in terms of factorial moments are
presented in Fig.~2.

\begin{figure}[htb]
\begin{center}
\mbox{\epsfig{file=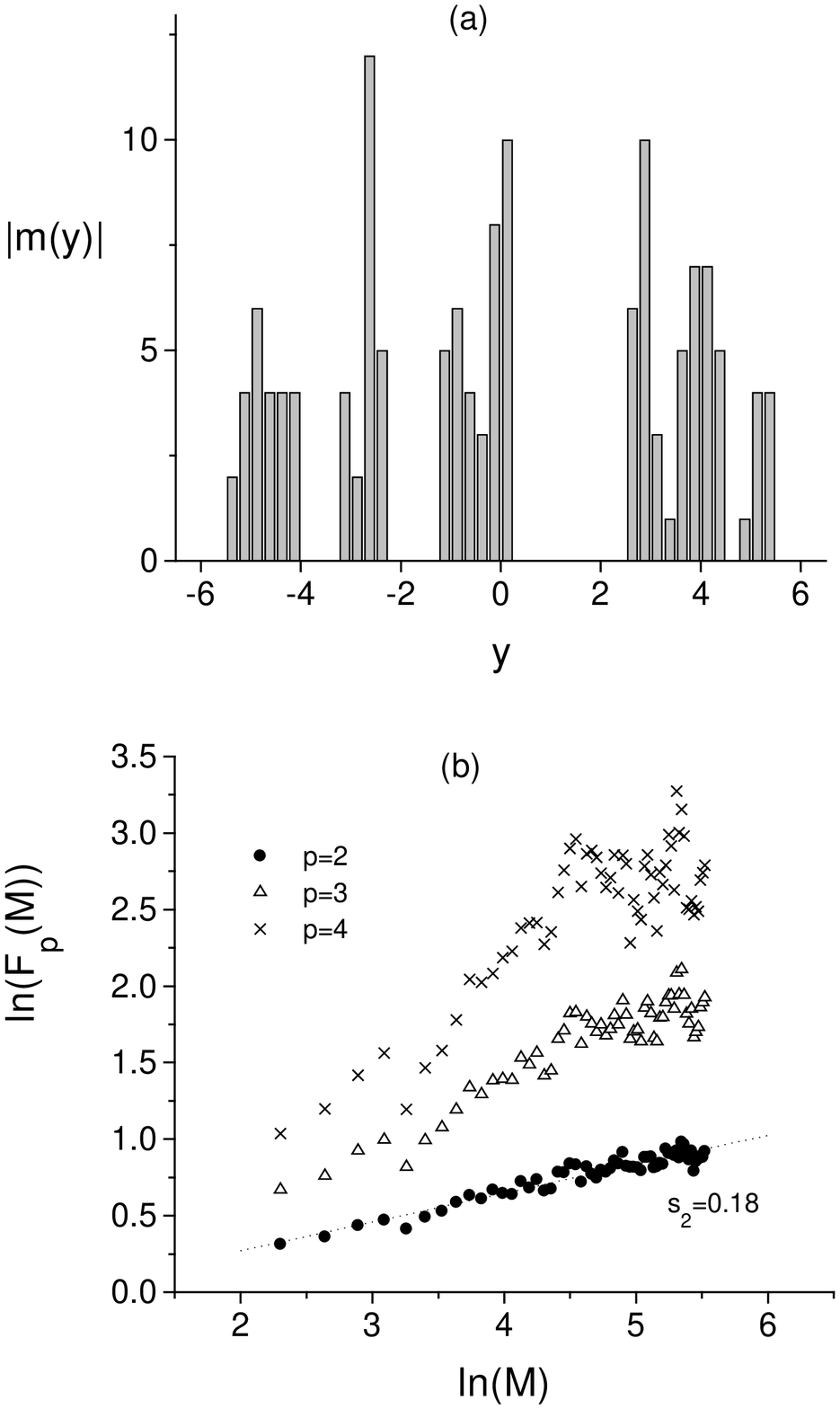,width=0.7\textwidth,angle=0}}
\caption{(a) The fluctuations of critical baryons in rapidity for a
MC-generated event.
(b) The first three factorial moments for the event shown in (a) in a
log-log plot. A linear fit determining the slope $s_2$ ($\approx 0.18$) of
the second moment is also shown.}
\label{fig:fi2}
\end{center}
\end{figure}

The intermittency exponent of the second moment $F_2$ in
rapidity is found to be $s_2 \approx 0.18$ which is very
close to the theoretically expected value ($\frac{1}{6}$) of a monofractal
$1-d$ set with fractal dimension $\frac{5}{6}$.
In conclusion, we have shown that measurements of net-baryon spectra in
rapidity provide a valuable set of obsevables in heavy ion experiments,
in connection with the phenomenology of the QCD critical point.

{}

\end{document}